\documentclass[reprint,amssymb,amsmath,superscriptaddress,aps,pra,nofootinbib,10pt]{revtex4-1}

\usepackage{color}
\usepackage{bm}
\usepackage{amsmath,amssymb}
\usepackage{amsfonts}
\usepackage{dsfont}
\usepackage{graphicx}
\usepackage{float}
\usepackage{subfigure}
\usepackage{verbatim}
\usepackage{bbold}
\usepackage{dcolumn}
\usepackage[dvipsnames]{xcolor}
\usepackage{listings}
\definecolor{blueprl}{RGB}{46,48,146}
\usepackage[pdftex,colorlinks=true,urlcolor=blueprl,citecolor=blueprl,linkcolor=blueprl]{hyperref}
\usepackage{mathrsfs}
\usepackage{filecontents}
\usepackage{mathtools}
\usepackage{times}

\newtheorem{proposition}{Proposition}

\newenvironment{proofb}[1][Proof]{\noindent\textbf{#1.} }{\ \rule{0.5em}{0.5em}}

\newcommand{\tr}{\mbox{tr}}

\begin{document}

\title{Maximum entanglement of formation for a two-mode Gaussian state over passive operations}
\author{Spyros Tserkis}  \email{spyrostserkis@gmail.com}
\affiliation{Centre for Quantum Computation and Communication Technology, Department of Quantum Science, Australian National University, Canberra, ACT 2601, Australia.}
\author{Jayne Thompson}
\affiliation{Centre for Quantum Technologies, National University of Singapore, 3 Science Drive 2, Singapore 117543, Republic of Singapore}
\affiliation{Horizon Quantum Computing, Alice@Mediapolis, 29 Media Circle, Singapore 138565, Republic of Singapore}
\author{Austin P. Lund}
\affiliation{Centre for Quantum Computation and Communication Technology, School of Mathematics and Physics, University of Queensland, St Lucia, Queensland 4072, Australia}
\author{Timothy C. Ralph}
\affiliation{Centre for Quantum Computation and Communication Technology, School of Mathematics and Physics, University of Queensland, St Lucia, Queensland 4072, Australia}
\author{Ping Koy Lam}
\affiliation{Centre for Quantum Computation and Communication Technology, Department of Quantum Science, Australian National University, Canberra, ACT 2601, Australia.}
\affiliation{Nanyang Quantum Hub, School of Physical and Mathematical Sciences, Nanyang Technological University, Singapore 639673, Republic of Singapore}
\author{Mile Gu} \email{mgu@quantumcomplexity.org}
\affiliation{Nanyang Quantum Hub, School of Physical and Mathematical Sciences, Nanyang Technological University, Singapore 639673, Republic of Singapore}
\affiliation{Complexity Institute, Nanyang Technological University, Singapore 639673, Republic of Singapore}
\author{Syed M. Assad} \email{cqtsma@gmail.com}
\affiliation{Centre for Quantum Computation and Communication Technology, Department of Quantum Science, Australian National University, Canberra, ACT 2601, Australia.}
\affiliation{Nanyang Quantum Hub, School of Physical and Mathematical Sciences, Nanyang Technological University, Singapore 639673, Republic of Singapore}

\begin{abstract}
We quantify the maximum amount of entanglement of formation (EoF) that can be achieved by continuous-variable states under passive operations, which we refer to as \textit{EoF-potential}. Focusing, in particular, on two-mode Gaussian states we derive analytical expressions for the EoF-potential for specific classes of states. For more general states, we demonstrate that this quantity can be upper-bounded by the minimum amount of squeezing needed to synthesize the Gaussian modes, a quantity called squeezing of formation. Our work, thus, provides a new link between non-classicality of quantum states and the non-classicality of correlations.
\end{abstract}

\maketitle

\section{Introduction}

\textit{Squeezed states} \cite{Walls.N.83} have been recognized as a non-classical resource that can be exploited to improve the performance of several tasks beyond their classical limit \cite{Caves.PRD.81,Braun.et.al.RMP.18,Pirandola.et.al.NP.18}, e.g., enhanced sensitivity of the LIGO detector \cite{Schnabel.et.al.NC.10,LIGO.NP.11,LIGO.NP.13}, and real-time phase-tracking \cite{Berry.Wiseman.PRA.02,Yonezawa.et.al.S.12}. The level of squeezing is characterized by the maximum amount the variance of a single mode falls below the vacuum limit, over all quadrature measurements. Recently, a measure to quantify the squeezing of a multi-mode state has been proposed, i.e., the \textit{squeezing of formation} (SoF) \cite{Idel.Lercher.Wolf.JPA.16}, which quantifies the minimum amount of squeezing needed to create a Gaussian state (spread over multiple modes).

Another fundamental resource in quantum information is \textit{entanglement} \cite{Einstein.Podolsky.Rosen.PR.35, SchrOdinger.MPC.89, Werner.PRL.89}. Entanglement is a property of quantum systems that is manifested as correlations of quantum observables that cannot be classically reproduced \cite{Masanes.Liang.Doherty.PRL.08}. This can lead to the violation of Bell inequalities \cite{Gisin.PLA.91}, steering \cite{Wiseman.Jones.Doherty.PRL.07}, or protocols such as quantum teleportation \cite{Pirandola.Mancini.LP.06, Cavalcanti.Skrzypczyk.Supic.PRL.17}, all of which are impossible classically. There is no unique quantifier of entanglement; three examples are entanglement of formation (EoF) \cite{Bennett.et.al.PRA.96}, logarithmic negativity (LogNeg) \cite{Vidal.Werner.PRA.02}, and relative entropy of entanglement (REE) \cite{Vedral.et.al.PRL.97}. Given that different entanglement measures are, in general, inequivalent to each other \cite{Virmani.Plenio.PLA.00}, the choice of a quantifier depends on the problem that is considered. A typical motivation for using LogNeg is that, unlike EoF or REE, for example, it is easy  to compute.

In the context of continuous-variable (CV) quantum information, the relationship between entanglement and squeezing has been discussed by several authors. Initial discussions appeared in Refs.~\cite{Scheel.Welsch.PRA.01, Kim.et.al.PRA.02} where it was conjectured that non-classicality (e.g., squeezing) is necessary for entanglement, which was later proven correct \cite{Xiang-bin.PRA.02}. In Ref.~\cite{Wolf.Eisert.Plenio.PRL.03}, Wolf, Eisert, and Plenio proved that under passive operations (assuming access to ancillary vacuum modes) any squeezed state can be transformed into an entangled state with a non-positive partial transpose with respect to a given partition. This result was recently improved to include all entangled states \cite{Lami.Serafini.Adesso.NJP.18}. The connection between non-classicality and entanglement has also been addressed under several other notions, e.g., fidelity \cite{Olivares.Paris.PRL.11}, Schmidt coefficients \cite{Vogel.Sperling.PRA.14}, nonclassical depth \cite{Brunelli.et.al.PRA.15}, quantum discord \cite{Brunelli.et.al.PRA.15, Fu.Luo.Zhang.EPL.19}, and nonclassicality invariants \cite{Arkhipov.et.al.PRA.16}.

The maximum entanglement generated using passive operations, as quantified by LogNeg, has been previous investigated -- and was shown to be analytically computable for any two-mode Gaussian state~\cite{Wolf.Eisert.Plenio.PRL.03}. In Ref.~\cite{Asboth.et.al.PRL.05} the concept of \textit{entanglement potential} was introduced (in the context of characterizing the nonclassicality of a single-mode state) as the maximum entanglement that can be produced through passive linear optics applied on a single-mode nonclassical state and ancillary vacuum modes, measured via LogNeg and REE, for a variety of nonclassical states.

In this work, we extend the analysis of Refs.~\cite{Wolf.Eisert.Plenio.PRL.03, Asboth.et.al.PRL.05}, by defining the entanglement potential of a given two-mode Gaussian state as the maximum attainable entanglement through passive linear optics, measured by EoF, and thus naming it \textit{EoF-potential}. For certain special cases, we derive closed form solutions for this quantity, and in the more general case, we show that the EoF-potential can be upper-bounded by the minimum amount of squeezing needed to synthesize the state (SoF).

This manuscript proceeds as follows. We begin in Sec.~\ref{secpre} with a brief review regarding Gaussian states, Gaussian operations, and the quantification of squeezing and entanglement. In Sec.~\ref{secmaxent} we define the notion of EoF-potential between two bosonic modes and demonstrate how it may be bounded from above by the SoF. In Sec.~\ref{seceofpotan} we then derive analytical expressions for the EoF-potential, while  Sec.~\ref{concl} concludes with avenues of further research.

\section{Preliminaries}
\label{secpre}

\subsection{Gaussian States}

Quantized bosonic modes, that are the main focus in this work, are described by CV states \cite{Weedbrook.et.al.RVP.12, Adesso.Ragy.OSID.14, Serafini.B.17, Holevo.B.19}. Those modes are associated with the quadrature field operators $\hat{x}_j:=\hat{a}_j+\hat{a}^{\dag}_j$ and $\hat{p}_j:=i(\hat{a}^{\dag}_j-\hat{a}_j)$, where $\hat{a}_j$ and $\hat{a}^{\dag}_j$ are the annihilation and creation operators, respectively, with $[\hat{a}_i,\hat{a}^{\dag}_j]=\delta_{ij}$, the kronecker delta.

Gaussian states are a subclass of CV states that can be fully described by the first two statistical moments (mean value and variance) of the quadrature field operators. For the purposes of this work we can ignore the mean value (fix it for simplicity to zero as it does not contribute to the squeezing or the entanglement of the state), and fully describe the $n$-mode Gaussian states through a real, symmetric and positive-definite matrix called the covariance matrix, whose elements are given by $\sigma_{ij}:=\frac{1}{2}\langle \{\hat{q}_{i},\hat{q}_{j}\}\rangle$, where $\hat{q}:=(\hat{x}_{1},\hat{p}_{1},\ldots, \hat{x}_{n},\hat{p}_{n})^{T}$ is the vectorial operator.

Every Gaussian state represented by a covariance matrix $\boldsymbol{\sigma}$ can be non-uniquely decomposed as \cite{Bhatia.B.07}
\begin{equation}
\boldsymbol{\sigma} = \boldsymbol{\pi} + \boldsymbol{\varphi} \,,
\label{gendec}
\end{equation}
where $\boldsymbol{\pi}$ is the covariance matrix of a pure Gaussian state, i.e., $\det (\boldsymbol{\pi})=1$, and $\boldsymbol{\varphi} \geqslant 0$ is a positive-semidefinite matrix representing random correlated displacements in the quadrature fields.

\subsection{Gaussian Unitary Operations}

The action of a Gaussian unitary operation (any unitary operation that transforms a Gaussian state into another Gaussian state) on the covariance matrix $\boldsymbol{\sigma}$ can be described by the following symplectic transformation $\Sigma$
\begin{equation}
\boldsymbol{\sigma} \mapsto \Sigma \boldsymbol{\sigma} \Sigma^T \,,
\end{equation}
where $\Sigma \Omega \Sigma^T= \Omega$, with $\Omega := \bigoplus_{i=1}^n  \begin{bmatrix}
0 & 1 \\
-1  & 0
\end{bmatrix}$, known as the symplectic form.

Quantum operations can be distinguished as \textit{passive} or \textit{active}, depending on whether they require an external source of energy to realize. From an operational point of view, passive operations are considered ``free'', since they can be straightforwardly implemented in the laboratory. On the other hand, active operations are typically more demanding.

\begin{figure}[!t]
\centering
\includegraphics[width=\columnwidth]{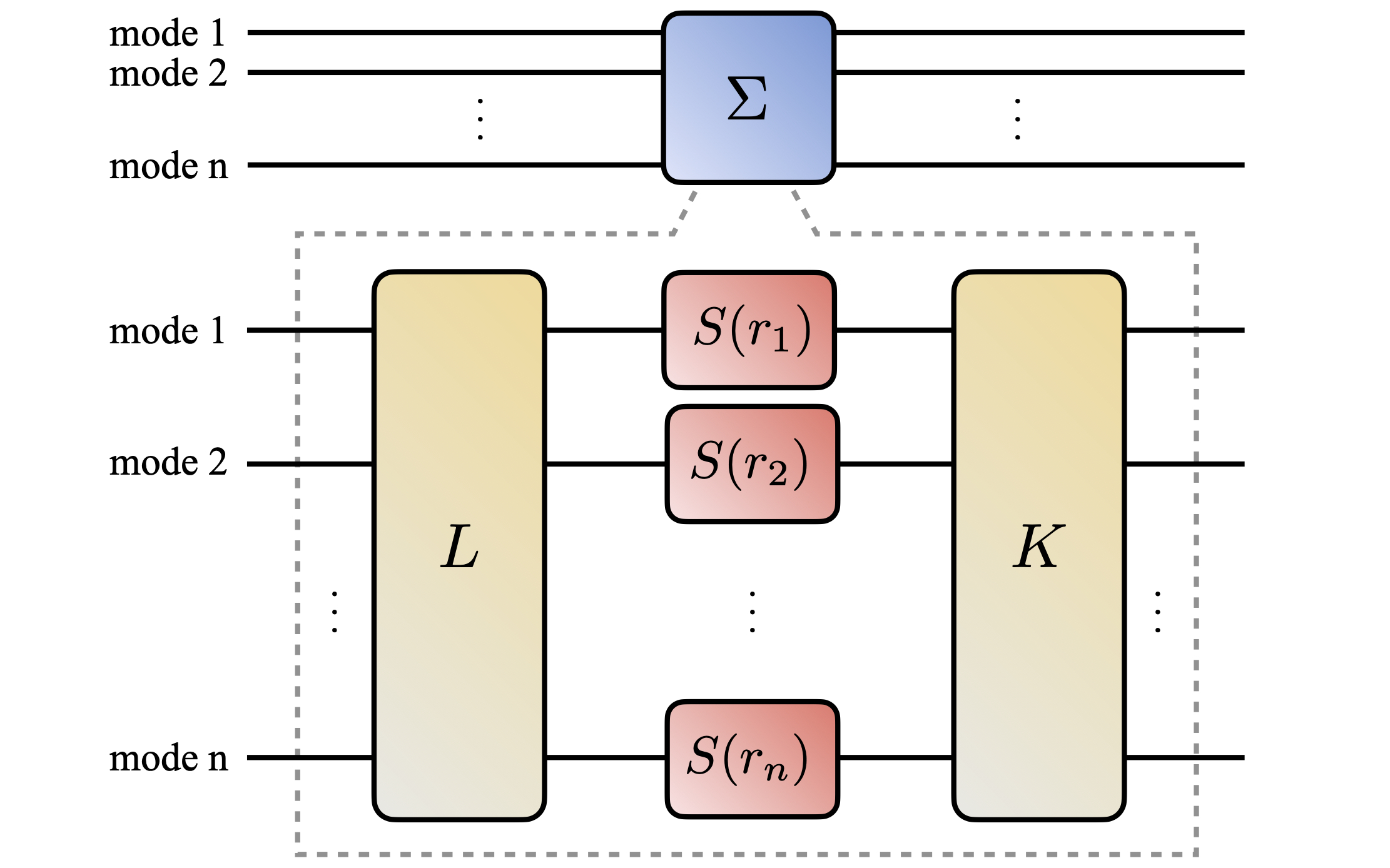}
\caption{Bloch-Messiah decomposition. A symplectic transformation $\Sigma$ can be uniquely decomposed into passive ($K,L$) and active ($S$) operations.}
\label{fig:figure1}
\end{figure}

Any symplectic operation $\Sigma$ can be decomposed through the \textit{Bloch-Messiah decomposition} \cite{Arvind.et.al.P.95,Braunstein.PRA.05}, into a sequence of passive and active operations (see Fig.~\ref{fig:figure1}) as follows
\begin{equation}
\Sigma = K \left[ \bigoplus_{i=1}^n S(r_i) \right] L \,,
\label{BlochMessiah}
\end{equation}
where $K$ and $L$ are symplectic passive operations, e.g., phase shifts or beam-splitters \cite{Reck.Zeilinger.Bernstein.PRL.94}, while $S(r_i)$ is a set of single-mode squeezing active operations, defined as
\begin{equation}
S(r_i):=\begin{bmatrix}
e^{r_i} & 0 \\
0  & e^{-r_i}
\end{bmatrix} \,,
\end{equation}
with $r_i \in \mathbb{R}$. The Bloch-Messiah decomposition will be used in the next section as an intuitive way for quantifying squeezing in pure Gaussian states.

\subsection{Squeezing of Formation}

Experimentally, squeezing is a sophisticated operation that involves non-linear processes, and thus practical limitations emerge in our ability to construct certain quantum states. For a single mode, the quantification of squeezing is straightforward, since we only have to check how much the variance of $\hat{x}$ or $\hat{p}$ beats the shot noise (the variance of any quadrature of a vacuum state), which can also be considered as a non-classicality measure \cite{Asboth.et.al.PRL.05, Brunelli.et.al.PRA.15, Arkhipov.et.al.PRA.16, Yadin.et.al.PRX.18, Kwon.et.al.PRL.19}. For multi-mode quantum states, though, where multiple interfering modes can be squeezed, the situation is more complicated, and defining a measure that reflects the ``total'' amount of squeezing in a state is a non-trivial task.

Given an $n$-mode pure Gaussian state $\boldsymbol{\pi}$, we can write $\boldsymbol{\pi}= \Sigma \Sigma^T$, and since $\Sigma$ has a non-unique Bloch-Messiah decomposition we can always choose a particular $\Sigma$ with $L=\mathds{1}$, i.e., $\boldsymbol{\pi}= K \left[ \bigoplus_{i=1}^n  S(r_i)  S^T(r_i) \right] K^T$ (see Fig.~\ref{fig:figure1}). Thus, applying the appropriate passive $n$-mode operation $K^T$, we end up with a covariance matrix in the following form
\begin{equation}
\boldsymbol{\pi} \mapsto K^T \boldsymbol{\pi} K = \bigoplus_{i=1}^n  S(r_i)  S^T(r_i) =\bigoplus_{i=1}^n \begin{bmatrix}
e^{2r_i} & 0 \\
0  & e^{-2r_i}
\end{bmatrix} \,.
\label{pure}
\end{equation}
This means that the pure state $\boldsymbol{\pi}$ can be decomposed into $n$ uncorrelated pure squeezed states through the passive operation $K^T$. The $2n$ eigenvalues $\lambda_{i}$ of this covariance matrix come in reciprocal pairs, so if we arrange them in an increasing order, i.e., $\lambda_i^{\uparrow} := \lambda_1 \leqslant \lambda_2 \leqslant \cdots \leqslant \lambda_{2n}$, we can fully characterize the eigenspectrum of $\boldsymbol{\pi}$ through the first half of them. The squeezing of this multi-mode pure state $\boldsymbol{\pi}$ can be characterized through the following function \cite{Idel.Lercher.Wolf.JPA.16}
\begin{equation}
\mathcal{S}(\boldsymbol{\pi}) := -\frac{1}{2} \sum_{i=1}^n \ln \left[\lambda_{i}^{\uparrow}(\boldsymbol{\pi}) \right] = \sum_{i=1}^n |r_i| \,,
\label{sofpure}
\end{equation}
which gives the sum of the absolute squeezing parameters of each mode for a pure state in the Bloch-Messiah decomposition.

\textit{Squeezing of formation} (SoF) for an arbitrary state $\boldsymbol{\sigma}$ is defined \cite{Idel.Lercher.Wolf.JPA.16} as the convex-roof extension of Eq.~(\ref{sofpure}), i.e.,
\begin{equation}
\mathcal{S}(\boldsymbol{\sigma}) := \inf_{\boldsymbol{\pi}} \left\{ \mathcal{S}(\boldsymbol{\pi}) \,\, \big| \,\, \boldsymbol{\sigma} = \boldsymbol{\pi} + \boldsymbol{\varphi}  \right\} \,,
\label{sof}
\end{equation}
and it quantifies the least amount of squeezing of the pure state $\boldsymbol{\pi}$ on which we can apply random correlated displacements $\boldsymbol{\varphi}$ to create the state $\boldsymbol{\sigma}$. The merit of this measure is that it rigorously characterizes the amount of squeezing in a state, since the function $\mathcal{S}(\boldsymbol{\sigma})$ satisfies properties such as convexity and continuity \cite{Idel.Lercher.Wolf.JPA.16}.

\paragraph*{Remark 1.} From the the Bloch-Messiah decomposition it follows that two states with the same SoF cannot in general be transformed under passive operations into one another. Consider for example two pure states with the same SoF. The symplectic operation applied onto the vacuum modes (for each state) has, in general, different local squeezing parameters $r_i$, and thus subsequent passive operations cannot make the states identical. For example, with current technology the highest amount of single mode squeezing has a parameter of $r \sim1.7$ (or $\sim$15 dB) \cite{Vahlbruch.et.al.PRL.16}. So, according to Eq.~(\ref{sofpure}), a given pure two-mode state of $\mathcal{S}(\boldsymbol{\sigma}) \sim 2$ can only be constructed if the squeezing is distributed such that $|r_1|+|r_2| \sim 2$, but each mode is not squeezed more than the (current) limit, such that $|r_i| \leqslant 1.7$.

\subsection{Entanglement of Formation}

\textit{Entanglement of formation} (EoF) is defined as the convex-roof extension of the entropy of entanglement \cite{Bennett.et.al.PRA.96}, and quantifies the entanglement in terms of the entropy of entanglement of the least entangled state needed to prepare it (under local operations and classical communication).

For two-mode Gaussian states EoF is given by \cite{Wolf.et.al.PRA.04, Ivan.Simon.arXiv.08, Marian.Marian.PRL.08, Tserkis.Ralph.PRA.17, Tserkis.Onoe.Ralph.PRA.19}
\begin{equation}
\mathcal{E} (\boldsymbol{\sigma}):= \inf_{\boldsymbol{\pi}} \left\{ \mathcal{E} (\boldsymbol{\pi})  \,\, \big| \,\, \boldsymbol{\sigma} = \boldsymbol{\pi} + \boldsymbol{\varphi} \right\} \,,
\label{eof}
\end{equation}
where $\mathcal{E} (\boldsymbol{\pi})$ is the \textit{entropy of entanglement} of the bipartite pure state $\boldsymbol{\pi}$. $\mathcal{E} (\boldsymbol{\pi})$ is defined as the von Neumann entropy of the reduced state, and given by \cite{Holevo.Sohma.Hirota.PRA.99,Adesso.Illuminati.PRA.05}
\begin{equation}
\mathcal{E}(\boldsymbol{\pi}):=\left\{
\begin{array}{cc}
      h \left[ \nu_- \left(\boldsymbol{\pi}^{\Gamma} \right) \right] & \text{for} \quad \nu_- \left(\boldsymbol{\pi}^{\Gamma} \right) < 1 \\
      0 & \text{for} \quad \nu_- \left(\boldsymbol{\pi}^{\Gamma} \right)  \geqslant 1 \\
\end{array}
\right. \,,
\label{eoe}
\end{equation}
where $\nu_- \left(\boldsymbol{\pi}^{\Gamma} \right)$ is the lowest (of the two) symplectic eigenvalue of the partially transposed pure state $\boldsymbol{\pi}$, and $h(x)$ is the auxiliary function defined as
\begin{equation}
h(x):=\frac{(1{+}x)^2}{4x} \log_2 \left[ \frac{(1{+}x)^2}{4x} \right]-\frac{(1{-}x)^2}{4x} \log_2  \left[ \frac{(1{-}x)^2}{4x} \right] \,.
\label{auxiliary}
\end{equation}

\paragraph*{Remark 2.} Eq.~(\ref{eof}) technically defines the Gaussian-EoF, which for multi-mode Gaussian states, in general, is an upper bound to the EoF, but for two-mode Gaussian states the two measures coincide \cite{Akbari-Kourbolagh.Alijanzadeh-Boura.QIP.15,Wilde.PRA.18}. For the rest of the work we focus only on two-mode Gaussian states.

\section{Entanglement Potential}
\label{secmaxent}

In this section, we introduce the concept of entanglement potential using EoF as the entanglement measure, and we present the main result of this work in Proposition 1, where we upper-bound the entanglement potential for any two-mode Gaussian state.

\subsection{Definition of EoF-potential}

\begin{figure}[!t]
\centering
\includegraphics[width=\columnwidth]{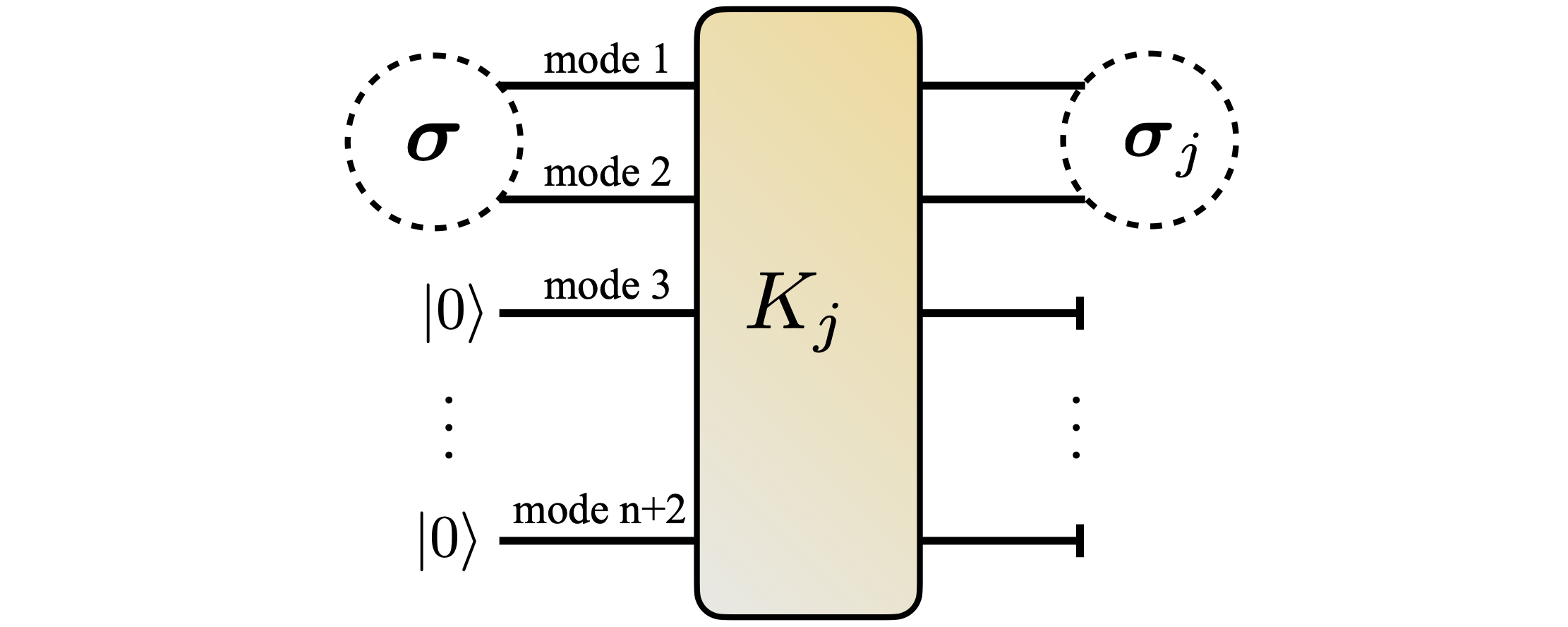}
\caption{Passive operations $K_j$ applied on a two-mode Gaussian state $\boldsymbol{\sigma}$ and $n$ ancillary vacuum modes. The final state $\boldsymbol{\sigma}_j$ is the reduced two-mode Gaussian state that remains after tracing out the last $n$ modes.}
\label{fig:figure2}
\end{figure}

Consider a two-mode Gaussian state with covariance matrix $\boldsymbol{\sigma}$, and access to $n$ ancillary vacuum modes and $(n{+}2)$-mode passive operations $K_j$. \textit{Entanglement of formation potential} (EoF-potential) quantifies the maximum attainable EoF of a state $\boldsymbol{\sigma}$ as follows
\begin{equation}
\mathcal{P} (\boldsymbol{\sigma}) := \sup_{j} \left\{  \mathcal{E} (\boldsymbol{\sigma}_{j})  \,\, \big| \,\, \boldsymbol{\sigma}_{j} =\tr_n \left[ K_j ( \boldsymbol{\sigma} \oplus \mathds{1}_n ) K_j^T \right] \right\} \,,
\label{peof}
\end{equation}
where $\tr_{n}$ denotes the partial trace over the last $n$ modes (see Fig.~\ref{fig:figure2}). The introduction of the ancillary vacuum modes in the definition of the EoF-potential might seem redundant at first sight, but an example where those extra modes can indeed be useful can be found in Sec.~\ref{symst}. Also, note that the definition given in Eq.~(\ref{peof}) for the EoF-potential can be modified for any other entanglement measure or monotone. For instance, based on this context the proposition 2 of Ref.~\cite{Wolf.Eisert.Plenio.PRL.03} corresponds to the LogNeg-potential.

\subsection{Bounds of EoF-potential}

By construction, the EoF of any state $\boldsymbol{\sigma}$ is a lower bound to the EoF-potential, i.e.,
\begin{equation}
\mathcal{E} (\boldsymbol{\sigma}) \leqslant \mathcal{P} (\boldsymbol{\sigma})  \,.
\label{lowbound}
\end{equation}

For two-mode Gaussian states we can always numerically estimate the value of $\mathcal{E} (\boldsymbol{\sigma})$, and for special classes we can also achieve analytical expressions \cite{Wolf.et.al.PRA.04,Ivan.Simon.arXiv.08,Marian.Marian.PRL.08,Tserkis.Ralph.PRA.17,Tserkis.Onoe.Ralph.PRA.19,Adesso.Illuminati.PRA.05}. A more interesting bound to the EoF-potential would be an upper bound, that we derive below.

\begin{proposition}
For an arbitrary two-mode Gaussian state $\boldsymbol{\sigma}$, the EoF-potential is upper-bounded as follows
\begin{equation}
\mathcal{P} (\boldsymbol{\sigma}) \leqslant h \left[ e^{- \mathcal{S}(\boldsymbol{\sigma})} \right] \,,
\label{upbound}
\end{equation}
where $\mathcal{S}(\boldsymbol{\sigma})$ is the SoF of $\sigma$.
\end{proposition}

\begin{proofb}
    The EoF-potential can be written as
    \begin{align}
    \mathcal{P} (\boldsymbol{\sigma}) :=& \sup_{j} \left\{  \mathcal{E} (\boldsymbol{\sigma}_{j})  \,\, \big| \,\, \boldsymbol{\sigma}_{j} =\tr_n \left[ K_j ( \boldsymbol{\sigma} \oplus \mathds{1}_n ) K_j^T \right] \right\} 	\nonumber \\
    =& \sup_{j} \inf_{\boldsymbol{\pi}_j} \left\{  \mathcal{E} (\boldsymbol{\pi}_{j})  \,\, \big| \,\, \boldsymbol{\sigma}_{j} = \boldsymbol{\pi}_{j} + \boldsymbol{\varphi}_{j} \right\} \nonumber \\
    =& \sup_{j} \mathcal{E} (\boldsymbol{\pi}_{e,j}) \,,
    \label{proofeq1}
    \end{align}
where $\boldsymbol{\pi}_{e,j}$ is the covariance matrix of the pure state that achieves the $\inf_{\boldsymbol{\pi}_j} \mathcal{E} (\boldsymbol{\pi}_{j})$. Based on Refs.~\cite{Wolf.Eisert.Plenio.PRL.03, Lami.Serafini.Adesso.NJP.18}, for any pure state $\boldsymbol{\pi}$ we have
\begin{equation}
\nu_- \left(\boldsymbol{\pi}^{\Gamma} \right) \geqslant \sqrt{\lambda_1^{\uparrow}(\boldsymbol{\pi}) \cdot \lambda_2^{\uparrow}(\boldsymbol{\pi})} \Rightarrow h \left[ \nu_- \left(\boldsymbol{\pi}^{\Gamma} \right) \right] \leqslant h \left[ e^{- \mathcal{S}(\boldsymbol{\pi})} \right],
\end{equation}
and for pure states we can always find a passive operation (without introducing ancillary vacuum modes) that saturates the above inequality, so for a generic pure state $\boldsymbol{\pi}$ we have
\begin{equation}
\sup_{j} \mathcal{E} (\boldsymbol{\pi}_j) = h \left[ e^{- \mathcal{S}(\boldsymbol{\pi}_j)} \right] \,,
\end{equation}
where the right-hand side does not depend on the operation $K_j$. Thus, the EoF-potential in Eq.~(\ref{proofeq1}) takes the following form
\begin{equation}
\mathcal{P} (\boldsymbol{\sigma}) = h \left[ e^{- \mathcal{S}(\boldsymbol{\pi}_{e,j})} \right] \,.
\label{peof2}
\end{equation}

A passive operation cannot affect the SoF of a state, so
\begin{equation}
\mathcal{S} (\boldsymbol{\sigma}_{j}) = \inf_{\boldsymbol{\pi}_j} \mathcal{S}(\boldsymbol{\pi}_{j}) = \mathcal{S} (\boldsymbol{\pi}_{s,j}) =\mathcal{S} (\boldsymbol{\sigma})  \,,
\end{equation}
where $\boldsymbol{\pi}_{s,j}$ is the covariance matrix of the state that achieves the $\inf_{\boldsymbol{\pi}_j} \mathcal{S} (\boldsymbol{\pi}_{j})$. For any passive operation $K_j$ we have
\begin{equation}
\mathcal{E}(\boldsymbol{\pi}_{e,j}) \leqslant \mathcal{E}(\boldsymbol{\pi}_{s,j}) \Rightarrow  \sup_j \mathcal{E} (\boldsymbol{\pi}_{e,j}) \leqslant \sup_j \mathcal{E} (\boldsymbol{\pi}_{s,j}) \,,
\end{equation}
and, so, we finally get
\begin{equation}
\mathcal{P} (\boldsymbol{\sigma}) \leqslant h \left[ e^{- \mathcal{S}(\boldsymbol{\sigma})} \right] \,,
\end{equation}
which completes the proof.
\end{proofb}

In order to calculate the upper bound in Eq.~(\ref{upbound}) we need to first calculate $\mathcal{S}(\boldsymbol{\sigma})$. In Ref.~\cite{Idel.Lercher.Wolf.JPA.16} a numerical method was derived for the estimation of $\mathcal{S}(\boldsymbol{\sigma})$, along with some upper and lower bounds. In the section below, we derive analytical expressions for some specific classes of two-mode Gaussian states.

\paragraph*{Remark 3.} Regardless of how each mode is individually squeezed, EoF-potential is related to the overall squeezing, quantified by SoF. Take for example a pure state $\boldsymbol{\pi}_1$ that consists of a position-quadrature squeezed vacuum ($r_1=2$) and a non-squeezed vacuum ($r_2=0$), and another state $\boldsymbol{\pi}_2$ that consists of two squeezed vacuum modes, one in the position quadrature ($r_1=1$) and the other in the momentum quadrature ($r_2=-1$). Those two states have the exact same amount of SoF, since based on Eq.~(\ref{sofpure}) we have $\mathcal{S}(\boldsymbol{\pi}_1)=|2|+|0|=2$ and $\mathcal{S}(\boldsymbol{\pi}_2)=|1|+|-1|=2$. The EoF for the two states is also equal, which can be seen as follows. We apply to both states a balanced (50:50) beam-splitter which (as a passive operation) keeps the SoF constant. Then, through local unitary operations (both passive and active), which keep the EoF constant, we can end up with the exact same covariance matrix (called the standard form of the covariance matrix \cite{Duan.et.al.PRL.00,Simon.PRL.00}), so $\mathcal{E}(\boldsymbol{\pi}_1)=\mathcal{E}(\boldsymbol{\pi}_2)$.

\section{Exact value for EoF-potential in special cases}
\label{seceofpotan}

In this section, we derive analytical expressions for the SoF of specific classes of two-mode Gaussian states, which in conjunction with the corresponding expressions for the EoF, give a closed formula for the EoF-potential.

Let us reorder the vectorial operator into $\hat{q}':=(\hat{x}_{1},\hat{x}_{2},\hat{p}_{1},\hat{p}_{2})^{T}$, so the covariance matrix of a two-mode Gaussian state takes the following form
\begin{equation}
\footnotesize
\boldsymbol{\sigma}= \begin{bmatrix}
\langle x_1^2 \rangle & \langle x_1 x_2 \rangle & \frac{1}{2}\langle \{ x_1, p_1 \} \rangle & \langle x_1 p_2 \rangle \\
\langle x_2 x_1 \rangle  & \langle x_2^2 \rangle & \langle x_2 p_1 \rangle & \frac{1}{2}\langle \{ x_2, p_2 \} \rangle \\
\frac{1}{2}\langle \{ p_1, x_1 \} \rangle & \langle p_1 x_2 \rangle & \langle p_1^2 \rangle & \langle p_1 p_2 \rangle \\
\langle p_2 x_1 \rangle &\frac{1}{2}\langle \{ p_2, x_2 \} \rangle & \langle p_2 p_1 \rangle & \langle p_2^2 \rangle\\
\end{bmatrix} \,.
\label{covmatrix}
\end{equation}

Consider a state with no cross quadrature correlations\footnote{This is not the most general class of states, since using only passive operations we cannot always end up to this type of covariance matrix, but it is an experimentally relevant class (see Ref.~\cite{Assad.et.al.arxiv.20} for a method to verify whether a state has no cross quadrature correlations has been constructed by some of us).}, i.e., $\langle \hat{x}_i \hat{p}_j \rangle=\langle \hat{p}_i \hat{x}_j \rangle=0 \,, \forall \ \{i,j \}$, so Eq.~(\ref{covmatrix}) becomes
\begin{equation}
\boldsymbol{\sigma}=
\begin{bmatrix}
a_1 & c_1\\
c_1 & b_1
\end{bmatrix} \oplus
\begin{bmatrix}
a_2 & c_2\\
c_2 & b_2
\end{bmatrix} =
\boldsymbol{c}_x \oplus \boldsymbol{c}_p  \,,
\label{covmatrix3}
\end{equation}
where we require that $a_i b_i - c_i\geqslant 1$ for the state to be physical. In order to simplify the calculation of SoF for states in the form of Eq.~(\ref{covmatrix3}) we show below that instead of minimizing over all pure states we can restrict the optimization over only a subset of them.

It has been shown in Refs.~\cite{Simon.Mukunda.Dutta.PRA.94,Wolf.et.al.PRA.04} that a two-mode Gaussian state is pure if and only if its covariance matrix has the following form
\begin{equation}
\boldsymbol{\pi}(\boldsymbol{z},\boldsymbol{y})=
\begin{bmatrix}
\boldsymbol{z} & \boldsymbol{z} \boldsymbol{y} \\
\boldsymbol{y} \boldsymbol{z} & \boldsymbol{y} \boldsymbol{z} \boldsymbol{y}+\boldsymbol{z}^{-1}
\end{bmatrix}  \,,
\label{purestate}
\end{equation}
where $\boldsymbol{z}=\boldsymbol{z}^T > 0$ and $\boldsymbol{y}=\boldsymbol{y}^T$ are real and symmetric $2 {\times} 2$ sub-matrices.

\begin{proposition}
Among all pure two-mode Gaussian states $\boldsymbol{\pi}(\boldsymbol{z},\boldsymbol{y})$, squeezing of formation is minimized by pure states of the form $\boldsymbol{\pi}(\boldsymbol{z},0)$, i.e.,
\begin{equation}
\mathcal{S}[\boldsymbol{\pi}(\boldsymbol{z},\boldsymbol{y})] \geqslant \mathcal{S}[\boldsymbol{\pi}(\boldsymbol{z},0)] \,.
\end{equation}
\end{proposition}

\begin{proofb}
Based on Eq.~(\ref{sofpure}), SoF for a state $\boldsymbol{\pi}(\boldsymbol{z},\boldsymbol{y})$ is equal or larger than that of a state $\boldsymbol{\pi}(\boldsymbol{z},0)$ if we have
\begin{equation}
\lambda_i^{\uparrow}[\boldsymbol{\pi}(\boldsymbol{z},\boldsymbol{y})] \leqslant \lambda_i^{\uparrow}[\boldsymbol{\pi}(\boldsymbol{z},0)] \,, \quad \text{for}\quad  i=\{1,2\} \,,
\end{equation}
where $\lambda_{i}^{\uparrow}[\boldsymbol{\pi}(\boldsymbol{z},\boldsymbol{y})]$ are the eigenvalues of each state in an increasing order. Then, for a given orthogonal projection $\Pi=(1,1,0,0)^T$ on $\boldsymbol{\pi}(\boldsymbol{z},\boldsymbol{y})$, i.e.,
\begin{equation}
\boldsymbol{z}=\Pi^T \boldsymbol{\pi} (\boldsymbol{z},\boldsymbol{y}) \Pi \,,
\end{equation}
the \textit{Cauchy interlacing theorem} (Theorem 4.3.17 in Ref.~\cite{Horn.Johnson.B.12}) imposes that
\begin{equation}
\lambda_i^{\uparrow}[\boldsymbol{\pi}(\boldsymbol{z},\boldsymbol{y})] \leqslant \lambda_i^{\uparrow}(\boldsymbol{z})=\lambda_i^{\uparrow}[\boldsymbol{\pi}(\boldsymbol{z},0)] \,,
\end{equation}
which completes the proof.
\end{proofb}

Based on the above proposition, for states in the form of Eq.~(\ref{covmatrix3}) the SoF needs to be minimized over only pure states of the form $\boldsymbol{\pi}(\boldsymbol{z},0)$, and due to Eq.~(\ref{gendec}) the constraint $\boldsymbol{\sigma} \geqslant \boldsymbol{\pi}$ becomes $\boldsymbol{c}_p^{-1}  \leqslant \boldsymbol{z} \leqslant \boldsymbol{c}_x$, i.e.,
\begin{equation}
\mathcal{S}[ \boldsymbol{\sigma}=\boldsymbol{c}_x \oplus \boldsymbol{c}_p ] := \inf_{\boldsymbol{z}}  \left\{ \mathcal{S} [\boldsymbol{\pi}(\boldsymbol{z},0)] \,\, \big| \,\, \boldsymbol{c}_p^{-1}  \leqslant \boldsymbol{z} \leqslant \boldsymbol{c}_x \right\} \,.
\label{sofx}
\end{equation}

Below, we present the analytical expressions for two specific classes of two-mode Gaussian states.

\subsection{Symmetric States}
\label{symst}

\textit{Symmetric states} have quadratures with equal variance, but (in general) different correlations in $\hat{x}$ and $\hat{p}$. The covariance matrix of symmetric states is given by
\begin{equation}
\boldsymbol{\sigma}_{\text{sym}}=
\begin{bmatrix}
a & c_1\\
c_1 & a
\end{bmatrix} \oplus
\begin{bmatrix}
a & c_2\\
c_2 & a
\end{bmatrix}  \,.
\end{equation}

Interfering the two modes on a balanced beam-splitter, we end up with the following covariance matrix
\begin{equation}
\begin{bmatrix}
a + c_1 & 0\\
0 & a - c_1
\end{bmatrix} \oplus
\begin{bmatrix}
a + c_2 & 0\\
0 & a - c_2
\end{bmatrix}  \,,
\end{equation}
which represents two uncorrelated modes, and the constraint $\boldsymbol{c}_p^{-1}  \leqslant \boldsymbol{z} \leqslant \boldsymbol{c}_x$ takes the form
\begin{equation}
\begin{bmatrix}
\frac{1}{a+c_2} & 0\\
0 & \frac{1}{a-c_2}
\end{bmatrix} \leqslant \boldsymbol{z} \leqslant
\begin{bmatrix}
a + c_1 & 0\\
0 & a - c_1
\end{bmatrix}  \,.
\end{equation}

Squeezing of formation, then, is equal to
\begin{equation}
\mathcal{S}(\boldsymbol{\sigma}_{\text{sym}})=- \ln (\mu_+ \mu_-) \,,
\label{SoFsym}
\end{equation}
where $\mu_{\pm}=\min \{ 1, \sqrt{a \pm c_1}, \sqrt{a \pm c_2}\}$. The EoF of symmetric states is given by \cite{Giedke.et.al.PRL.03}
\begin{equation}
\mathcal{E} (\boldsymbol{\sigma}_{\text{sym}}) =\left\{
\begin{array}{cc}
      h \left( \sqrt{\lambda_1^{\uparrow} \lambda_2^{\uparrow}} \right) & \text{for} \quad \lambda_1^{\uparrow} \lambda_2^{\uparrow} < 1 \\
      0 & \text{for} \quad \lambda_1^{\uparrow} \lambda_2^{\uparrow}  \geqslant 1 \\
\end{array}
\right. \,,
\end{equation}
where $\lambda_1^{\uparrow}$ and $\lambda_2^{\uparrow}$ are the two lowest eigenvalues of $\boldsymbol{\sigma}_{\text{sym}}$ in an increasing order, i.e., $\lambda_1^{\uparrow} \leqslant \lambda_2^{\uparrow}$. When the state $\boldsymbol{\sigma}_{\text{sym}}$ is classical, i.e., $\lambda_1^{\uparrow} \geqslant 1$, or entangled with $\lambda_2^{\uparrow} < 1$, we get
\begin{equation}
\mathcal{E} (\boldsymbol{\sigma}_{\text{sym}}) =h \left[ e^{- \mathcal{S}(\boldsymbol{\sigma}_{\text{sym}})} \right]\,,
\label{equality_of_bound}
\end{equation}
and thus the lower bound in Eq.~(\ref{lowbound}) coincides with the upper bound in Eq.~(\ref{upbound}). However, for states (separable or entangled) with $\lambda_1^{\uparrow} < 1 \leqslant \lambda_2^{\uparrow}$, we get $\mathcal{E} (\boldsymbol{\sigma}_{\text{sym}}) < h \left[ e^{- \mathcal{S}(\boldsymbol{\sigma}_{\text{sym}})} \right]$ which does not saturate the upper bound. In order to achieve the upper bound, we can perform the following operation
\begin{equation}
	\boldsymbol{\sigma}' =\tr_3 \left[ K ( \boldsymbol{\sigma}_{\text{sym}} \oplus \mathds{1} ) K^T \right] \,,
	\label{operation_potential}
\end{equation}
with $K=[B(\pi/4)\oplus \mathds{1}] [ \mathds{1} \oplus B(\pi/2)] [B(\pi/4)\oplus \mathds{1}]$. $B(\theta)=\exp[\theta(\hat{a}^{\dag}\hat{b}-\hat{a}\hat{b}^{\dag})]$ is a beam-splitter operation, with $\hat{a}$ and $\hat{b}$ denoting the annihilation operators of each mode respectively. Practically, in Eq.~(\ref{operation_potential}), we first apply a balanced beam-splitter on the first two modes in order to diagonalize the covariance matrix, corresponding to two uncorrelated modes, one thermal and one squeezed. Without loosing generality we assume that the thermal mode is the second one. Then, the ancilla vacuum is swapped with the thermal mode, and a second balanced beam-splitter is applied to increase the entanglement. Finally, the operation $\tr_3$ traces out the 3rd mode. At the end, the entanglement of the state $\boldsymbol{\sigma}'$ is given by 
\begin{equation}
\mathcal{E} (\boldsymbol{\sigma}') =h \left[ e^{- \mathcal{S}(\boldsymbol{\sigma}_{\text{sym}})} \right]\,,
\end{equation}
which coincides with the upper bound [Eq.~(\ref{upbound})], and thus the EoF-potential of symmetric states is given by
\begin{equation}
\mathcal{P} (\boldsymbol{\sigma}_{\text{sym}}) = h \left( \mu_+ \mu_- \right) \,.
\end{equation}

\subsection{Balanced Correlated States}
\label{balcorst}

\textit{Balanced correlated states} have an equal amount of correlations in $\hat{x}$ and $\hat{p}$, i.e., $c_1=-c_2=c>0$, as well as quadratures with the same variance in $\hat{x}$ and $\hat{p}$, i.e., $a_1=a_2=a$, and $b_1=b_2=b$. Their covariance matrix has the following form
\begin{equation}
\boldsymbol{\sigma}_{\text{bc}}=
\begin{bmatrix}
a & c\\
c & b
\end{bmatrix} \oplus
\begin{bmatrix}
a & -c\\
-c & b
\end{bmatrix}  \,.
\label{bcstate}
\end{equation}

These states correspond to the output of a two-mode squeezed state passing through two independent phase-invariant Gaussian channels, and they are typically encountered in quantum communication protocols, e.g., quantum teleportation \cite{Pirandola.Mancini.LP.06}, and quantum key distribution \cite{Pirandola.et.al.arxiv.19}. If $1-a-b+a b \geqslant c^2$ all the eigenvalues of $\boldsymbol{\sigma}_{\text{bc}}$ are greater or equal to one, and thus the state is classical and SoF vanishes.

Unlike symmetric states, these states cannot be brought into a diagonal form by a passive transformation. In App.~\ref{app1}, we show that in this case the optimal $\boldsymbol{z}$ is given by
\begin{equation}
\boldsymbol{z}_{\text{opt}} =\frac{\lambda_+}{2}
\begin{bmatrix}
1&1\\
1&1
\end{bmatrix}
+\frac{ \lambda_-}{2}
\begin{bmatrix}
1&-1\\
-1&1
\end{bmatrix} \,,
\end{equation}
where $ \lambda_-$ and $\lambda_+$ are the two eigenvalues
of $\boldsymbol{z}_{\text{opt}}$, given by
\begin{equation}
  \lambda_+ = \frac{1}{\lambda_-}= \frac{a+b+2c}{1+a b -c^2 +\sqrt{(1-a
    b+c^2)^2 -(a-b)^2}} \,,
\label{lambda}
\end{equation}
and the SoF is
\begin{equation}
\mathcal{S}(\boldsymbol{\sigma}_{\text{bc}})= \ln \lambda_+ \,.
\label{SoFbc}
\end{equation}

The EoF for balanced correlated states has been calculated for the first time in Ref.~\cite{Adesso.Illuminati.PRA.05} (where the name ``GMEM" was given to the same type of states), but a simpler expression can be found in Refs.~\cite{Tserkis.Ralph.PRA.17,Tserkis.Onoe.Ralph.PRA.19}. As we show in App.~\ref{app2}, the EoF is equal to
\begin{equation}
\mathcal{E} (\boldsymbol{\sigma}_{\text{bc}}) = h \left[ e^{- \mathcal{S}(\boldsymbol{\sigma}_{\text{bc}})} \right] = h( \lambda_-) \,,
\label{equality}
\end{equation}
and, since the lower bound in Eq.~(\ref{lowbound}) coincides with the upper bound in Eq.~(\ref{upbound}), the EoF-potential for balanced correlated states is given by
\begin{equation}
\mathcal{P} (\boldsymbol{\sigma}_{\text{bc}}) = h( \lambda_-) \,.
\end{equation}
Thus, the entanglement of any balanced correlated state achieves by construction its EoF-potential, and no further passive operation is needed.

\section{Discussion}
\label{concl}

In continuous-variable optics, the class of operations that is typically considered the most simple to realize is passive linear operations. In this work we ask, given two-mode Gaussian states, what is the maximum amount of entanglement (as quantified by entanglement of formation) that we can synthesize using such operations? We named this quantity EoF-potential, and analytically computed it for certain special cases of two-mode Gaussian states. For more general states, we demonstrated that EoF-potential can be bounded from above given the squeezing of formation of the same state -- the amount of squeezing needed to synthesize these modes. This, thus, presents  another interesting connection between non-classicality of quantum states and the non-classicality of correlations in the Gaussian regime.

There are a number of interesting future directions. One immediate question is the tightness of the bounds derived; another being to what extent such relations can be generalized to other measures of quantum correlations and non-classicality of quantum states. This would become especially pertinent should we consider non-Gaussian states, whereby existing measures of non-classicality have already been defined from the perspectives of metrological advantage and non-equilibrium energy~\cite{Yadin.et.al.PRX.18,Kwon.et.al.PRL.19,Narasimhachar.et.al.arXiv.19}. Such directions could well shed light to potential connections to entanglement distillation; where a no-go theorem prevents distillation through local Gaussian operations \cite{Eisert.Scheel.Plenio.PRL.02,Fiurasek.PRL.02,Giedke.Cirac.PRA.02}.

\paragraph*{Note added:}Recently, we learnt about a similar paper \cite{Hertz.Cerf.DeBievre.PRA.20}, where the authors bound EoF using a non-classicality measure called \textit{monotone of total noise}.

\section{Acknowledgements}

We thank H. Jeng for useful discussions. This work is supported by the Australian Research Council (ARC) under the Centre of Excellence for Quantum Computation and Communication Technology (Grant No. CE170100012), the Singapore Ministry of Education Tier 1 grant 2019-T1-002-015 (RG162/19), Singapore National Research Foundation Fellowship NRF-NRFF2016-02 and the NRF-ANR grant NRF2017-NRF-ANR004 VanQuTe.

Any opinions, findings and conclusions or recommendations expressed in this material are those of the author(s) and do not reflect the views of National Research Foundation, Singapore.

\appendix

\section{SoF for Balanced Correlated States}
\label{app1}

In this appendix, we compute the SoF for a balanced correlated state [defined in Eq.~(\ref{bcstate})] through Eq.~(\ref{sofx}). Following the analysis of Ref.~\cite{Wolf.et.al.PRA.04}, we set
\begin{equation}
  \boldsymbol{c}_x :=\begin{bmatrix}x_0+x_1&x_2\\x_2&x_0-x_1\end{bmatrix} \,, \quad
  \boldsymbol{c}_p^{-1} :=\begin{bmatrix}p_0+p_1&p_2\\p_2&p_0-p_1\end{bmatrix} \,,
 \end{equation}
and
\begin{equation}
 \boldsymbol{z} :=\begin{bmatrix}z_0+z_1&z_2\\z_2&z_0-z_1\end{bmatrix} \,,
 \end{equation}
and we represent the matrices $\boldsymbol{z}$, $\boldsymbol{c}_x $, and $\boldsymbol{c}_p^{-1}$, as points in three dimensions with coordinates, e.g. $(x_1,x_2,x_0)$. With this representation, and given the constraint $\boldsymbol{c}_p^{-1}  \leqslant \boldsymbol{z} \leqslant \boldsymbol{c}_x$, the coordinates of $\boldsymbol{z}$ must lie within the intersection of two 45-degrees vertical cones with vertices corresponding to $\boldsymbol{c}_x$ and $\boldsymbol{c}_p^{-1}$. For states where both $\boldsymbol{c}_x$ and $\boldsymbol{c}_p$ have one eigenvalue greater than one and one eigenvalue less than one ($\boldsymbol{\sigma}_\text{bc}$ falls in this category), the optimal $\boldsymbol{z}$ must also lie on the surface of both these cones. This surface is parameterized by $\det( \boldsymbol{z}-\boldsymbol{c}_x)=0$ and $\det( \boldsymbol{c}_p^{-1} -\boldsymbol{z})=0$. With these two constraints, the problem of quantifying the SoF reduces to finding the optimal $\boldsymbol{z}$, restricted to the ellipse defined by the intersection of the two surfaces of the cones.

The coordinates of this ellipse can be parametrized by the angle $\phi$ as follows
 \begin{subequations}
 \begin{gather}
   \label{eq:12}
   z_1(\phi) = \frac{x_1+p_1}{2} + r_1 \cos \gamma \sin \phi -
                 r_2 \sin \gamma \cos \phi \,, \\
   z_2(\phi) = \frac{x_2+p_2}{2} + r_1 \sin \gamma \sin \phi +
                 r_2 \cos \gamma \cos \phi \,, \\
   z_0(\phi) = x_0 -\sqrt{(x_1-z_1)^2 + (x_2-z_2)^2} \,,
 \end{gather}
  \end{subequations}
where $\gamma := \arctan (c_2/c_1)$, $c_i := (a_i - b_i)/2 $, $r_1 := c_0$, and $r_2:=\sqrt{c_0^2-(c_1^2+c_2^2)}$. The SoF of $\boldsymbol{z}$ is $\mathcal{S} [\boldsymbol{\pi}(\boldsymbol{z},0)]=\frac{1}{2}\log \frac{\lambda_+}{\lambda_-}$, where
\begin{equation}
  \label{eq:14}
  \lambda_\pm = z_0 \pm \sqrt{z_1^2+z_2^2}= z_0\left(1 \pm \sqrt{\frac{z_1^2 + z_2^2}{z_0^2}}\right) \,,
\end{equation}
are the two eigenvalues of $\boldsymbol{z}$. Minimizing $\mathcal{S} [\boldsymbol{\pi}(\boldsymbol{z},0)]$ is equivalent to minimizing $(z_1^2 +z_2^2)/z_0^2$, so, we need to find the angle $\phi$ that minimizes $(z_1^2+z_2^2)/z_0^2$. For an arbitrary $\boldsymbol{c}_x$ and $\boldsymbol{c}_p^{-1}$, this involves solving a transcendental equation which does not have an analytical solution. However, for the balance correlated state, this minimization can be done analytically, and through long but straightforward computations, we find that  $(z_1^2 +z_2^2)/z_0^2$ is minimized when
\begin{equation}
z_0 =\frac{ \lambda_+ + \lambda_-}{2} \,, \quad
z_1 =\frac{ \lambda_- - \lambda_+}{2} \,, \quad \text{and} \quad
z_2 =0 \,,
\end{equation}
with
\begin{equation}
  \lambda_+ = \frac{1}{\lambda_-}=\frac{a+b+2c}{1+a\,b-c^2+\sqrt{(1-a\,b+c^2)^2-(a-b)^2}}\, .
\end{equation}

\onecolumngrid
\section{Proof of Eq.~(\ref{equality})}
\label{app2}

Based on Refs.~\cite{Tserkis.Ralph.PRA.17,Tserkis.Onoe.Ralph.PRA.19} the EoF for entangled balanced correlated states is given by
\begin{equation}
	\mathcal{E}(\boldsymbol{\sigma}_{\text{bc}})=\cosh^2r_o \log _2\left(\cosh ^2r_o\right)-\sinh ^2 r_o \log _2\left(\sinh ^2 r_o\right) \,,
\end{equation}
where
\begin{equation}
	r_o=\frac{1}{4} \ln \left[\frac{2 \kappa^2 + \tau_1 \tau_2 -2 \sqrt{\kappa^2 \left(\kappa^2+\tau_1 \tau_2\right)}}{\tau_2^2}\right]  \,,
\end{equation}
with
\begin{subequations}
\begin{align}
	\kappa &:=a b-c^2+1 \,, \\
	\tau_1 &:=2 c+a+b \,, \\
	\tau_2 &:=2 c-a-b \,.
\end{align}
\end{subequations}
The parameter $\lambda_-$ in Eq.~(\ref{lambda}) can analogously be written as
\begin{equation}
    \lambda_-= \frac{\kappa+\sqrt{\kappa^2 + \tau_1 \tau_2}}{\tau_1}  \,.
\end{equation}
In order to prove Eq.~(\ref{equality}) we need to show that
\begin{equation}
      \cosh^2 r_o = \frac{(1+e^{-2 r_o})^2}{4 e^{ -2r_o}} =\frac{(1+\lambda_-)^2}{4 \lambda_-} \quad \Longrightarrow \quad e^{-2r_o}=\lambda_- \quad \Longrightarrow \quad e^{-4r_o}=\lambda_-^2\,.
\end{equation}
Consider
\begin{subequations}
\begin{align}
	\lambda^2_- e^{4r_o} &= \frac{\Big[2 \kappa^2+ \tau_1 \tau_2 -2 \sqrt{\kappa^2 \left(\kappa^2+\tau_1 \tau_2\right)} \Big] \Big[ 2\kappa^2+\tau_1 \tau_2 +2\kappa\sqrt{\kappa^2+\tau_1 \tau_2} \Big]}{\tau_1^2 \tau_2^2}  \\
	&= \frac{(2\kappa^2 +\tau_1 \tau_2)^2 - 4\kappa^2(\kappa^2+\tau_1 \tau_2)}{\tau_1^2 \tau_2^2} \label{subequa} \\
	&= 1 \,.
\end{align}
\end{subequations}
Eq.~(\ref{subequa}) holds only when $\kappa \geqslant 0$, which is true for all quantum states, which completes the proof.

\twocolumngrid

\end{document}